\documentclass[sigconf]{acmart}

\copyrightyear{2026}
\acmYear{2026}
\setcopyright{cc}
\setcctype{by}
\acmConference[Q-SE '26]{17th International Workshop on Quantum Software Engineering}{April 12--18, 2026}{Rio de Janeiro, Brazil}
\acmBooktitle{17th International Workshop on Quantum Software Engineering (Q-SE '26), April 12--18, 2026, Rio de Janeiro, Brazil}
\acmDOI{10.1145/3786150.3788613}
\acmISBN{979-8-4007-2383-4/2026/04}

\usepackage{bo-macros}
\usepackage{amsthm}
\usepackage{graphicx}
\usepackage{subcaption}
\usepackage{balance}
\usepackage{enumitem}
\usepackage[braket, qm]{qcircuit}

\usepackage[defaultlines=3,all]{nowidow}
\newcommand{\sys}{\textsc{Twirlator}\xspace}

\begin{document}

\title{\sys: A Pipeline for Analyzing Subgroup Symmetry Effects in Quantum Machine Learning Ansatzes}

\author{Valter Uotila}
\email{first.last@aalto.fi}
\affiliation{%
  \institution{Aalto University \& University of Helsinki}
  \city{Espoo}
  \country{Finland}
}

\author{Väinö Mehtola}
\email{first.last@vtt.fi}
\affiliation{%
  \institution{VTT}
  \city{Espoo}
  \country{Finland}
}

\author{Ilmo Salmenperä}
\email{first.last@helsinki.fi}
\affiliation{%
  \institution{University of Helsinki}
  \city{Helsinki}
  \country{Finland}
}

\author{Bo Zhao}
\email{first.last@aalto.fi}
\affiliation{%
  \institution{Aalto University}
  \city{Espoo}
  \country{Finland}
}

\renewcommand{\shortauthors}{Uotila et al.}

\begin{abstract}
Symmetry is a strong inductive bias in geometric deep learning and its quantum counterpart, and has attracted increasing attention for improving the trainability of QML models. Yet incorporating symmetries into quantum machine learning (QML) ansatzes is not free: symmetrization often adds gates and constrains the circuits. To understand these effects, we present \textbf{\sys}, which is an automated pipeline that symmetrizes parameterized QML ansatzes and quantifies the trade-offs as the amount of symmetry increases. \sys models partial symmetries by the size of a subgroup of the symmetric group, enabling analysis between the ``no symmetry'' and ``full symmetry'' extremes. Across 19 common ansatz patterns, \sys symmetrizes circuits with respect to any subgroup of $S_n$ and measures (1) generator drift, (2) circuit overhead (depth and size), and (3) expressibility and entangling capability. The experimental evaluation focuses on subgroups of $S_4$ and $S_5$. \sys reveals that larger subgroups typically increase circuit overhead, reduce expressibility, and often increase entangling capability. The pipeline and results provide practical guidance for selecting ansatz patterns and symmetry levels that balance hardware cost and model performance in symmetry-aware QML applications.
\end{abstract}


\maketitle

\section{Introduction}


Many foundational classical machine learning models benefit from additional structure imposed by the learning tasks. Convolutional models exploit translation equivariance in vision~\cite{LeCun_Boser_Denker_Henderson_Howard_Hubbard_Jackel_1989,Lecun_Bottou_Bengio_Haffner_1998}, and graph neural networks enforce permutation symmetry for relational data~\cite{Kipf_Welling_2017,Scarselli_Gori_Tsoi_Hagenbuchner_Monfardini_2009}. Beyond conceptual elegance, symmetry constraints can produce concrete gains in accuracy and data efficiency. For instance, group-equivariant CNNs reduce test error on rotated MNIST from $5.03\%$ (standard CNN) to $2.28\%$ (rotation-equivariant P4CNN) under matched settings~\cite{cohen2016gcnn}. In scientific machine learning, equivariant architectures can also reduce data requirements by orders of magnitude: NequIP~\cite{tan2025highperformancetraininginferencedeep} reports competitive accuracy with 1,000$\times$ fewer training samples in a molecular modeling benchmark~\cite{batzner2022nequip}. These successes underpin geometric deep learning~\cite{Bronstein_Bruna_Cohen_Velickovic_2021} and have recently carried over to \emph{geometric} and \emph{equivariant} quantum machine learning (QML)~\cite{Meyer_Mularski_Gil_Fuster_Mele_Arzani_Wilms_Eisert_2023,Ragone_Braccia_Nguyen_Schatzki_Coles_Sauvage_Larocca_Cerezo_2023,Nguyen_Schatzki_Braccia_Ragone_Coles_Sauvage_Larocca_Cerezo_2024,Schatzki_Larocca_Nguyen_Sauvage_Cerezo_2024,Nha_Minh_Le_Kiss_Schuhmacher_Tavernelli_Tacchino_2025}, where symmetry-aware models can possibly improve trainability, generalization and avoid certain problems such as barren plateaus~\cite{West_Heredge_Sevior_Usman_2024,McClean_Boixo_Smelyanskiy_Babbush_Neven_2018,Holmes_Sharma_Cerezo_Coles_2022}.


From a \emph{quantum software engineering} perspective, ``using symmetry'' is a pipeline decision that should be identified, engineered, validated, and benchmarked. Incorporating symmetries into QML models affects the QML software stack end-to-end: after identifying symmetries in the learning task, they influence circuit construction, compilation/transpilation, the trainability and learnability of QML models, and resource estimations under hardware constraints. Concretely, stronger symmetry can turn a circuit that is deployable under a depth/size budget into one that is not. The current tooling provides limited reusable automation for applying symmetry transformations consistently across different ansatzes and learning tasks and for producing metric-driven analysis that makes these trade-offs explicit to practitioners.

Although many learning tasks have natural underlying symmetries, especially in physics~\cite{Mehtola_2025}, there is no quantum software that would enable us to symmetrize quantum machine learning ansatzes in an automated manner. For example, Qiskit, Pennylane, and Cirq provide limited tools for symmetrizing QML models. As noted in~\cite{Ragone_Braccia_Nguyen_Schatzki_Coles_Sauvage_Larocca_Cerezo_2023}, the practical realization of these models requires a solid understanding of group representation theory. 

Symmetry is often treated as a binary choice (none vs.\ full symmetry), while \emph{partial} symmetry---which may provide a better cost--performance balance---is rarely explored systematically~\cite{Ragone_Braccia_Nguyen_Schatzki_Coles_Sauvage_Larocca_Cerezo_2023}. To quantify the impact of increasing symmetry, we approach the symmetrization in terms of subgroups. While most of the previous research has not explicitly formulated symmetrization in terms of subgroups, we present this straightforward generalization. Subgroup-based symmetrization can be considered to produce partially equivariant models, which have been more extensively studied in classical geometric and equivariant machine learning~\cite{10.5555/3600270.3602912,10.5555/3540261.3542560,10.5555/3737916.3740994}. Nevertheless, this study shows that employing subgroup-based symmetries rather than the whole group produces more expressive and shorter ansatzes.

\sys structures the evaluation around three research questions:
\textbf{(RQ1)} How does subgroup symmetrization alter ansatz generators as symmetry increases?
\textbf{(RQ2)} How does circuit overhead (depth and gate count) scale with subgroup size across common ansatzes?
\textbf{(RQ3)} How do expressibility and entangling capability change as symmetry increases? 

\noindent This paper makes following contributions:

\tinyskip
\mypar{(1) A reusable symmetry-aware QML pipeline}
Based on angle encoding, induced unitary representations of symmetry groups, and the Pauli twirling formula, we implement an automated pipeline that symmetrizes some of the most common quantum machine learning ansatz patterns with respect to the subgroups. \sys automatically generalizes to any common ansatzes.

\tinyskip
\mypar{(2) Subgroup-based symmetry as a tunable knob in the QML model construction}
We propose that symmetry can be understood as a tunable knob in QML models, since one need not use the entire group but can incorporate only subgroups, as this work demonstrates. This reframes symmetry as a continuum rather than a binary choice, thereby supporting the systematic exploration of symmetry effects.

\tinyskip
\mypar{(3) A benchmark-driven characterization of symmetry}
We concretely study the effects of the gate symmetrization process by computing a difference between original and symmetric generators, which we call \emph{generator drift}. Additionally, we obtain circuit-related metrics, such as depth, and the expressibility and entangling capabilities of the symmetrized ansatzes for varying degrees of symmetry. We have open sourced \sys\footnote{\url{https://github.com/valterUo/twirlator}}.


\section{Background}\label{sec:background}


\sys treats symmetry as an explicit engineering knob in QML ansatz design. This section introduces the background needed to operationalize that knob: (i) how we model task symmetries with subgroups of a symmetric group, (ii) how \sys enforces a target subgroup via the twirling formula, and (iii) the expressibility and entangling capability metrics used in our evaluation.

\subsection{Symmetries in learning tasks}

Symmetries can appear in (quantum) machine learning in multiple ways~\cite{Bronstein_Bruna_Cohen_Velickovic_2021, Meyer_Mularski_Gil_Fuster_Mele_Arzani_Wilms_Eisert_2023, Mehtola_2025}. Symmetries can be discrete or continuous, and they are often divided into equivariant or invariant symmetries~\cite{Mehtola_2025}. Symmetries can appear only in the source data, or we can identify that the mapping, i.e., the learning problem, respects certain symmetries due to the problem's nature. In this work, we focus on symmetries that appear in the learning problem.

In this work, we assume an input space $\mathcal{X}$ and an output space $\mathcal{Y}$ for a given machine learning problem~\cite{Shalev_Shwartz_Ben_David_2014}. Given a set of samples $\mathcal{S} = \left\{ (x_i, y_i) \in \mathcal{X} \times \mathcal{Y} \right\}_{i=1}^{N}$, the goal is to learn a function $f \colon \mathcal{X} \to \mathcal{Y}$ so that it approximates the unknown distribution $D \subset \mathcal{X} \times \mathcal{Y}$. Depending on the learning task, it is possible to identify that the distribution $D \subset \mathcal{X} \times \mathcal{Y}$ respects certain symmetries, which can be encoded in the model.

In this work, symmetries are modeled using symmetric groups $S_n$ and their subgroups $S'_k \subset S_n$. The connection between the group theory and the vector spaces (data) is given by representation theory~\cite{Ragone_Braccia_Nguyen_Schatzki_Coles_Sauvage_Larocca_Cerezo_2023}. We can define a representation of a group $S_n$ on a vector space $V$ as a mapping $\varphi \colon S_n \times V \to V$, which satisfies that $\varphi(s) \colon V \to V$ is linear for every $s \in S_n$, $\varphi(e, v) = v$ for the neutral element $e \in S_n$ and $\varphi(s_1, \varphi(s_2, v)) = \varphi(s_1s_2, v)$ for every $s_1, s_2 \in S_n$ and $v \in V$. Using this representation, we obtain $\varphi(S'_k)$, which is the corresponding subgroup of linear mappings $V \to V$. 

Next, we assume that $\mathcal{X} \subset V$. Then, the mapping $f$ is invariant under the symmetric subgroup $S'_k$ if $f(\varphi(s, x)) = f(x)$ for all $x \in \mathcal{X}$ and for all $s \in S'_k$. With this definition, the level of symmetry in the mapping can be quantified by considering the size of the subgroup $S'_k \subset S_n$, for which the invariance holds. If the subgroup is a trivial one-element group consisting of only the neutral element, the data does not admit any symmetries. Neutral element is always mapped to the identity matrix $I$ and this does not change the vector, so that $f(I[x]) = f(x)$ is trivially satisfied. On the other hand, if the subgroup satisfies $S'_k = S_n$, the data is maximally symmetric with respect to the fixed symmetry group $S_n$.

In this work, we consider that the representation $\varphi \colon S_n \times V \to V$ is also realized by unitary matrices such that $\phi(S_n) = \left\{ U_s \mid s \in S_n \right\}$. Let $U_{\mathrm{init}}$ be the initial data encoding layer for the quantum machine learning model as a unitary operator. We consider special \textit{induced unitary representations of symmetry groups} which are required to satisfy the following condition
\begin{equation}\label{eq:induced_unitary_representation}
    U_{\mathrm{init}}(\varphi(s, x)) = U_s  U_{\mathrm{init}}(x) U_s^{\dag},
\end{equation}
for all $x \in \mathcal{X}$. This condition necessarily encodes the fact that the induced representation $\left\{ U_s \mid s \in S_n \right\}$ in a certain sense commutes with the data encoding layer $U_{\mathrm{init}}(x)$ with respect to the group $S_n$.

Now the symmetry in the learning problem is modeled with unitaries $\left\{U_s \mid s \in S_n \right\}$ which also satisfy \autoref{eq:induced_unitary_representation}. The equation establishes the connection between data encoding and symmetries. It is also important to note that \autoref{eq:induced_unitary_representation} is defined at the operator level, not the state level. 

\subsection{Gate symmetrization}

This work implements gate symmetrization, which is performed with the Pauli twirling formula~\cite{Meyer_Mularski_Gil_Fuster_Mele_Arzani_Wilms_Eisert_2023}. For this subsection, we assume that for a fixed data encoding $U_{\mathrm{init}}$ and a subgroup $S_k'$ of the symmetry group $S_n$, we have computed the induced unitary representations as defined in \autoref{eq:induced_unitary_representation}. 

Since the construction involves parameterized quantum circuits, the symmetrization relies on generators. All of the gates in the ansatzes can be expressed in terms of fixed generators as $R_{G}(\theta) = e^{-i\theta G}$, where $G$ is the generator for the parametrized gate $R_{G}(\theta)$.  Based on Proposition 1 (Commuting generators)~\cite{Meyer_Mularski_Gil_Fuster_Mele_Arzani_Wilms_Eisert_2023}, it suffices to apply the symmetrization process only to the generators. Hence, assume a fixed gate set expressed in terms of generators. Then, we define the Pauli twirling formula~\cite{Meyer_Mularski_Gil_Fuster_Mele_Arzani_Wilms_Eisert_2023, Helsen_2019} over the subgroup $S_k'$:
\begin{equation}\label{eq:twirling}
    \mathcal{T}[G] = \frac{1}{|S'_k|} \sum_{s \in S'_k}U_sGU_s^{\dag}.
\end{equation}
Applying the twirling formula to the generators $\mathcal{G}$ of the original gate set, we obtain the equivariant generator set $\mathcal{T}[\mathcal{G}]$. These generators are used to construct the corresponding symmetric parameterized circuit.

\subsection{Expressibility}

Expressibility of a parametrized quantum circuit refers to the circuit's ability to express pure states so that they cover the Hilbert space~\cite{https://doi.org/10.1002/qute.201900070}. In practice, the expressibility is computed based on fidelities which are estimated so that we sample pairs of states $|\varphi_{\theta}\rangle$ and $|\varphi_{\phi}\rangle$ for different randomly initialized parametrizations $\theta$ and $\phi$. Then, the corresponding fidelities $F = |\langle \varphi_{\theta} | \varphi_{\phi}\rangle|^2$ are considered as random variables. Let $P(F, \theta)$ be the estimated probability distribution of fidelities for parameters $\theta$, which results from sampling states from the parametrized circuit. Let $P_{\mathrm{Haar}}(F)$ be the Haard distribution~\cite{Mele2024introductiontohaar}, which can be computed with the closed form formula $P_{\mathrm{Haar}}(F) = (N-1)(1-F)^{N-2}$. The distributions $P(F, \theta)$ and $P_{\mathrm{Haar}}(F)$ are compared with Kullback-Leibler (KL) divergence, i.e., relative entropy~\cite{Kullback_Leibler_1951}
\begin{displaymath}
    D_{\mathrm{KL}}(P(F, \theta) \| P_{\mathrm{Haar}}(F)) = \sum_x P(F, \theta)[x] \log \frac{P(F, \theta)[x]}{P_{\mathrm{Haar}}[x]}.
\end{displaymath}
The expressibility of a parametrized quantum circuit is defined as its KL-divergence.

\subsection{Entangling capability}

The entangling capability of a parameterized quantum circuit quantifies how close, on average, the state it produces is to the maximally entangled state. Following the description in~\cite{https://doi.org/10.1002/qute.201900070}, we have computed Meyer-Wallach (MW) entanglement measure~\cite{Meyer_Wallach_2002}. Let $\Theta = \left\{ \theta_i \mid 1\leq i\leq n\right\}$ be a collection of parameter vectors for a parameterized quantum circuit. Previous work~\cite{Azad_Sinha_2023} defines Meyer-Wallach entanglement measure as
\begin{displaymath}
Q = \frac{2}{|\Theta|}\sum_{\theta_{i} \in \Theta}
            \left(1-\frac{1}{n}\sum_{k=1}^{n}\mathrm{Tr}(\rho_{k}^{2}(\theta_{i}))\right),
\end{displaymath}
where $n$ is the number of qubits and $\rho_k$ is the reduced density matrix of qubit $k$. The value $\mathrm{Tr}(\rho_{k}^{2})$ is the purity of the qubit $k$. If qubit $k$ is entangled with others, then the state $\rho_k$ becomes mixed and $\mathrm{Tr}(\rho_{k}^{2}) < 1$. Hence, this formula can be used to estimate the total entangling capability of the parameterized circuit. According to this measure, if a parameterized circuit produces only separable (product) states, its entangling capability is $0$. In contrast, a circuit that creates highly entangled states attains a value close to $1$. Previous research~\cite{https://doi.org/10.1002/qute.201900070} identified that the entangling capability of Haar random states is around $0.82$.
\section{Methodology}\label{sec:methodology}

Using the symmetry knob and metrics defined in \S\ref{sec:background}, we now describe \sys's evaluation methodology. The methodology consists of three steps: (i) selecting a representative collection of ansatzes~\cite{https://doi.org/10.1002/qute.201900070} and enumerating subgroups of $S_4$, (ii) generating subgroup-symmetrized circuits via the twirling formula, and (iii) comparing original and symmetrized circuits using generator drift, circuit overhead, expressibility, and entangling capability metrics.

\subsection{Selected ansatzes and subgroups}

\begin{figure}
    \centering
    \input{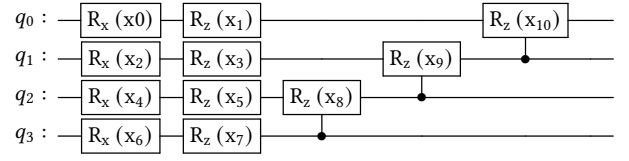}
    \caption{Ansatz $3$ from~\cite{https://doi.org/10.1002/qute.201900070} with four qubits}
    \label{fig:circuit_example_original}
    \Description{A quantum circuit diagram for a four-qubit variational ansatz. The circuit consists of an initial layer of parameterized single-qubit rotations (RX and RZ gates) on each qubit, followed by a layer of controlled-RZ (CRZ) entangling gates connecting adjacent qubits in a linear chain (q3 to q2, q2 to q1, and q1 to q0). All gates feature a total of eleven trainable parameters.}
\end{figure}

\sys studies various symmetry-related properties of 19 common ansatz patterns from~\cite{https://doi.org/10.1002/qute.201900070}. The ansatze implementations used in this work originate from~\cite{jern2025fine}, which provides them as PennyLane circuits. An example ansatz with id $3$ is presented in Fig.~\ref{fig:circuit_example_original}. Making the results partially comparable to~\cite{https://doi.org/10.1002/qute.201900070}, we focus on four-qubit versions of these ansatzes. Because of selecting four qubits, we construct the symmetric group $S_4$ and consider its possible subgroups, which have orders as $1$ (trivial one-element subgroup), $2$, $3$, $4$, $6$, $8$, $12$, and $24$ (complete $S_4$ has $4! = 24$ elements). We have considered all of the subgroups of $S_4$.

\subsection{Subgroup-based symmetrization}

\begin{figure*}
    \centering
    \input{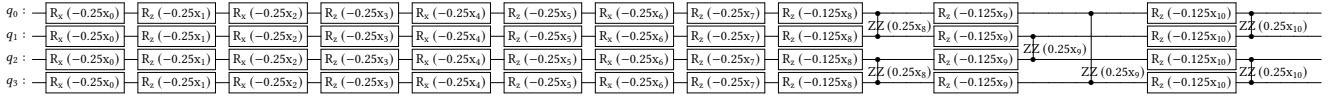}
    \caption{Symmetrized circuit corresponding the circuit in Fig.~\ref{fig:circuit_example_original} using a four element subgroup of symmetric group $S_4$}
    \label{fig:circuit_example_twirled}
    \Description{A quantum circuit diagram representing the symmetrized version of the Ansatz 3 circuit from Fig. 1, using a four-element subgroup of the symmetric group S4. The circuit consists of three main blocks applied sequentially across four qubits (q0 to q3): a repeated four-fold RX rotation layer (gates x0 to x7), a layer of ZZ entangling gates with shared parameters (x8 and x9), and a final layer combining RX and controlled ZZ gates (CZZ) with a shared parameter (x10).}
\end{figure*}

Since most of the gates in the ansatzes are parametrized, the system relies on Proposition 1 about commuting generators~\cite{Meyer_Mularski_Gil_Fuster_Mele_Arzani_Wilms_Eisert_2023}, which states that it suffices to symmetrize the generators instead of parametrized gates as unitaries. To compute the generators for the gates in the ansatzes, we use PennyLane's \texttt{qml.generator} function. Then, the system applies the previously described induced unitary representations as permutation matrices to the generators. Next, it uses the Pauli twirling formula in~\autoref{eq:twirling} to obtain the equivariant set of symmetrized generators. Finally, the same set of parameters is assigned for each symmetrized generator, and the resulting set of generators is synthesized into parametrized quantum circuits using Qiskit. When this process is applied to the example circuit in Fig.~\ref{fig:circuit_example_original} using a fixed four-element subgroup of the symmetric group $S_4$, we obtain a circuit in Fig.~\ref{fig:circuit_example_twirled}.

\subsection{Evaluation metrics}

Next, we introduce the evaluation metrics used to assess performance differences between the original ansatzes and their symmetrized counterparts. Let $A_i$ be the ansatz from~\cite{https://doi.org/10.1002/qute.201900070} for $1\leq i \leq 19$. Let $\mathcal{G}(A_i)$ be the set of generators for the gates in the ansatz. Let $\mathcal{T}[\mathcal{G}(A_i)]$ be the equivariant gateset for the gates defined by generators in $\mathcal{G}(A_i)$, when they are symmetrized with respect to a fixed subgroup. In the following, the norms are Frobenius norms. Then, we consider the three classes of metrics as follows.

\tinyskip
\mypar{Generator drift}
To quantify the difference between the original ansatzes and their symmetrized versions, we compute the average difference of the norm 
\begin{displaymath}
|| G - G_{\mathrm{twirl}}||,
\end{displaymath}
for $G_{\mathrm{twirl}} \in \mathcal{T}[\mathcal{G}(A_i)]$ and $G \in \mathcal{G}(A_i)$, where $G_{\mathrm{twirl}}$ is the symmetrized version of $G$. In other words, this means that the average norm for the projection onto the symmetric subspace is defined as
    \begin{equation}\label{eq:operator_difference_norm}
        \frac{1}{|\mathcal{T}[\mathcal{G}(A_i)]|}\sum_{G_{\mathrm{twirl}} \in \mathcal{T}[\mathcal{G}(A_i)]}\frac{1}{|\mathcal{G}(A_i)|} \sum_{G \in \mathcal{G}(A_i)} || G - G_{\mathrm{twirl}} ||.
    \end{equation}
Note that the previous value is interestingly depth-invariant. The value is computed for each gate's generator and then averaged over all gates in the ansatz. Since increasing the number of layers in the ansatz structures repeats the same gates multiple times, the value does not depend on the ansatz depth. Hence, in the results section, we present the results without referencing the depth.

\tinyskip
\mypar{Circuit-related metrics} When the gates in ansatz $A_i$ are replaced with the corresponding symmetrized gates based on generators in $\mathcal{T}_{\mathcal{U}}[\mathcal{G}(A_i)]$, we synthesize the circuit using the same gates as the original ansatzes have. Synthesis is performed using the Qiskit transpile function with the highest optimization level $3$. The results slightly depend on the synthesis method, and Qiskit does not perform all possible optimizations because the synthesized circuits contain parameters. For example, one can identify that some $R_z$ gates could still be combined in the circuit in Fig.~\ref{fig:circuit_example_twirled}, reducing the gate count and depth slightly. Nevertheless, these metrics provide a concrete quantification of the overhead introduced by symmetrization as a function of the system's degree of symmetry. Circuit-related metrics also enable us to compare those ansatzes that are expensive to symmetrize in terms of gate overhead. We observed that the overall trends for the total number of gates, two-qubit gates, and circuit depth are consistent. Therefore, we present the results using absolute gate counts in terms of the subgroup size.

\tinyskip
\mypar{Expressibility and entangling capability} For the same symmetrized circuits as in the previous point, we compute the expressibility and entangling capability as defined in~\cite{https://doi.org/10.1002/qute.201900070}. The values are calculated with open-source QLeet software~\cite{Azad_Sinha_2023}, which we updated~\footnote{\url{https://github.com/valterUo/qleet-light}}. We use the hyperparameters from~\cite{https://doi.org/10.1002/qute.201900070} so that we sample $10\ 000$ states for which we compute the independent fidelities (i.e., the same state does not appear twice in the fidelity computations). Although QLeet allows continuous expressibility comparison, we follow the description in~\cite{https://doi.org/10.1002/qute.201900070} and approximate the distributions $P(F, \theta)$ and $P_{\mathrm{Haar}}(F)$ with $75$ bins. With this setup, we obtained results that were closely similar to those of the previous research~\cite{https://doi.org/10.1002/qute.201900070} for the original ansatzes without symmetrization. We also used $10 \ 000$ states to compute the entangling capability. This sample size was found statistically robust~\cite{https://doi.org/10.1002/qute.201900070}.

\section{Implementation}\label{sec:implementation}

\sys implements subgroup-based ansatz symmetrization for parameterized circuits under discrete permutation symmetries. It takes as input a circuit and a subgroup, and outputs a symmetrized circuit along with artifacts such as induced representations and symmetrized generators.

\tinyskip
\mypar{Precomputed subgroups} 
Using the precomputed subgroups for $S_4$, we construct the unitary representations that satisfy the induced unitary representation definition in \autoref{eq:induced_unitary_representation}. For a fixed symmetric group $S_n$ and its subgroup $S'_k \subset S_n$, the elements in the group are permutations. Since we focus on angle encoding, which is expressed in terms of a diagonal unitary $U_{\mathrm{init}}$, this means that the induced representations $U_s$ are simply permutation matrices, which are easy to construct. If we introduce entanglement into the data-encoding unitary $U_{\mathrm{init}}$ and apply, for example, amplitude encoding, the resulting representations are more complex to build. Considering induced unitary representations for advanced data encoding methods is part of future research.

The system also contains other precomputed symmetric group structures for extended experiments. Using Sage Math, we have sampled at most $30$ random subgroups of size $k$ for each $1 \leq k \leq |S_n|$ for $3 \leq n \leq 9$. Some experiments also rely on the corresponding five-qubit circuits and the group $S_5$. Generally, the implementation handles any parametrized circuit with respect to any subgroup of a symmetric group, provided the model uses angle encoding.

\tinyskip
\mypar{Induced unitary representations}
Based on the permutations in the subgroup, the corresponding permutation matrix, i.e., the induced unitary representation, is constructed as follows. Let $n$ be the number of qubits in the system and $\sigma$ be a permutation on $\{0, 1, \dots, n-1\}$ such that $\sigma(i)$ gives us the target position of qubit $i$. This means we want to encode an action that moves the bit originally at position $i$ to position $\sigma(i)$. We define $\sigma^{-1}$ as the inverse permutation satisfying $\sigma^{-1}(\sigma(i)) = i$. The corresponding qubit permutation unitary $U_{\sigma} \in \mathbb{C}^{2^n \times 2^n}$ operates on computational basis states as
\begin{displaymath}
U_{\sigma} \, |b_0 b_1 \dots b_{n-1}\rangle = |b_{\sigma^{-1}(0)} \, b_{\sigma^{-1}(1)} \, \dots \, b_{\sigma^{-1}(n-1)}\rangle.
\end{displaymath}
The bit originally at position $i$ moves to position $\sigma(i)$.

\section{Evaluation}
\label{sec:evaluation}

Computationally, the most expensive experiments are those for expressibility and entangling capability, since they require simulating circuits. For those, we utilized 760 CPUs divided into 95 tasks. Each task was allocated 8 CPUs with 4GB of memory, and the running time ranged from 30 minutes to 2 hours, depending on the depth of the circuits. The software used for quantum computing simulations was Qiskit, and Pennylane was used for ansatz construction and generator computation.

\subsection{RQ1: How much does symmetrization change the generators?}

\begin{figure*}[t]
	\centering
	\begin{subfigure}[t]{0.4\textwidth}
		\centering
		\includegraphics[width=0.95\linewidth]{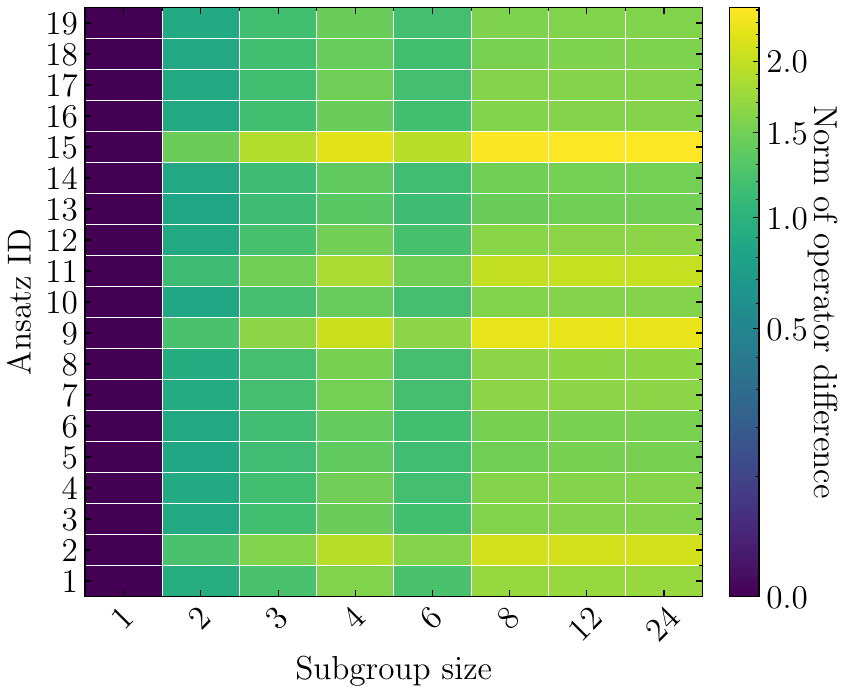}
		\caption{$S_4$ (four qubits).}
		\label{fig:commutator_norm_with_twirling}
	\end{subfigure}
	\hfill
	\begin{subfigure}[t]{0.4\textwidth}
		\centering
		\includegraphics[width=0.95\linewidth]{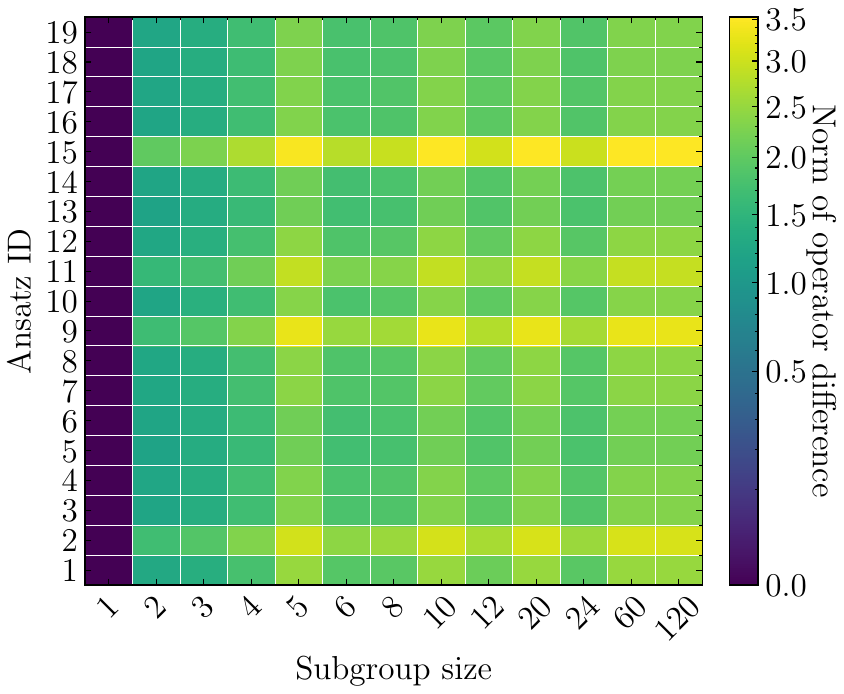}
		\caption{$S_5$ (five qubits).}
		\label{fig:commutator_norm_with_twirling_5}
	\end{subfigure}
	\caption{Generator drift under subgroup-based symmetrization.
		We report the operator difference norm (\autoref{eq:operator_difference_norm}) between original gate generators and their subgroup-twirled counterparts as symmetry strength increases (subgroup size on the x-axis).
		Higher values indicate larger generator changes induced by symmetrization.}
	\label{fig:generator_drift_heatmaps}
	\Description{Two heatmaps reporting generator drift (operator difference norm) across 19 ansatz patterns (y-axis) and increasing subgroup sizes (x-axis). Left: four-qubit ansatzes under subgroups of $S_4$ (orders 1,2,3,4,6,8,12,24). Right: five-qubit ansatzes under subgroups of $S_5$ (orders 1,2,3,4,5,6,8,10,12,20,24,60,120). Colors indicate the magnitude of generator drift; subgroup size 1 yields zero drift and larger subgroups generally increase drift.}
\end{figure*}

\tinyskip
\mypar{Four qubits ($S_4$)} The results for the operator difference norm, i.e., generator drift, which we defined in~\autoref{eq:operator_difference_norm}, are presented in Fig.~\ref{fig:commutator_norm_with_twirling} for four-qubit ansatzes. When the subgroup size is $1$, meaning that the subgroup consists of only the group's neutral element, we correctly obtain $0$ as a difference since $G = G_{\mathrm{twirl}}$ holds always. The norm increases with the size of the symmetry group, as a larger number of symmetries in the problem necessitates more extensive modifications to the generator $G$. The average generator drift across ansatzes increases from $0$ (subgroup size $1$) to $1.72$ (subgroup size $24$). Some interesting ansatzes are $2$, $9$, $11$ and $15$, whose generator drifts are $2.11$, $2.26$, $2.01$, and $2.42$ respectively when the subgroup size is $24$, and these are visible as yellow stripes. The commonality between these ansatzes is that they all contain non-parametrized gates, such as CNOTs, CZs, and Hadamard gates, and we suspect that their symmetrization slightly increases the difference. We can also see that the norm for subgroups of size $6$ is slightly lower than for subgroups of sizes $4$ and $8$, which makes this case stand out.

\tinyskip
\mypar{Five qubits ($S_5$)} A similar trend is observed for larger instances. Corresponding results for the 5-qubit ansatzes computed over the subgroups of $S_5$ are presented in Fig.~\ref{fig:commutator_norm_with_twirling_5}. In these cases, the average generator drift across ansatzes increases from $0$ to $2.51$. Ansatzes $2$, $9$, $11$, and $15$ again perform differently. We also observe that subgroups of sizes $5$, $10$, $20$, $60$, and $120$ show higher norms. Looking at the subgroups more closely, one can identify that many subgroups of size $5$ are contained in subgroups of size $10$, many of size $10$ are contained in subgroups of size $20$, and so on. Subgroups that contain smaller subgroups consequently inherit their properties. This would explain the stripes in Fig.~\ref{fig:commutator_norm_with_twirling} and Fig.~\ref{fig:commutator_norm_with_twirling_5}. 

\subsection{RQ2: How much circuit overhead does symmetrization introduce?}

\begin{figure*}[t]
    \centering
    \begin{subfigure}[t]{0.45\textwidth}
        \includegraphics[width=\textwidth]{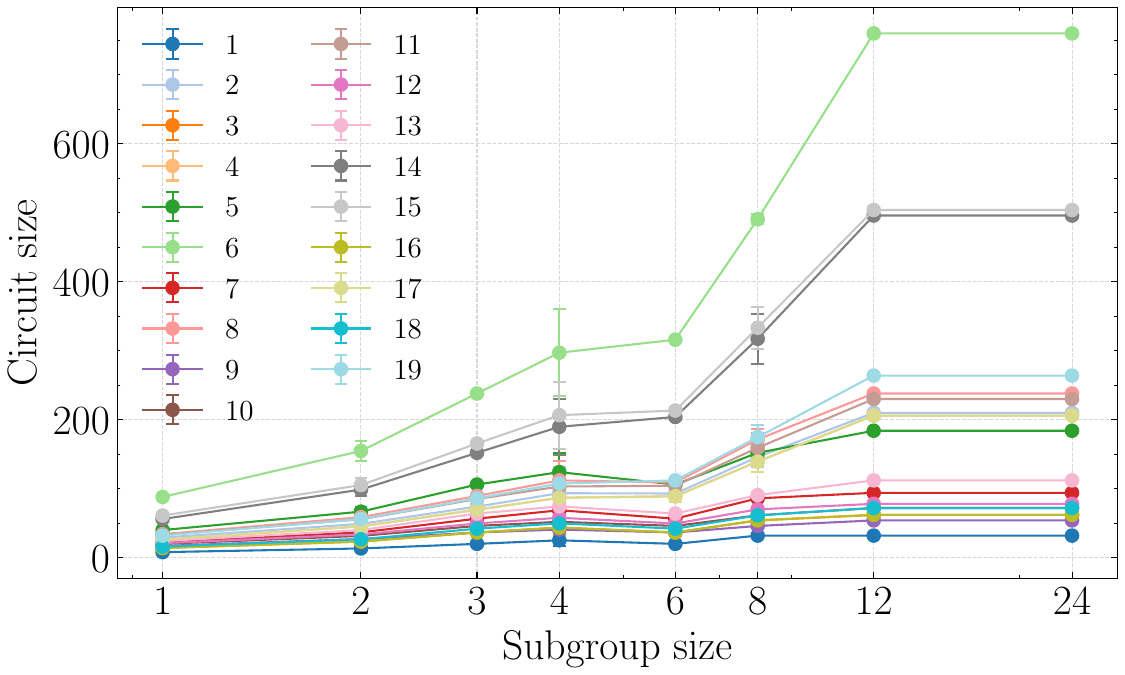}
        \caption{$S_4$ (four qubits, depth $1$).}
        \label{fig:absolute_circuit_sizes}
    \end{subfigure}
    \hfill
    \begin{subfigure}[t]{0.49\textwidth}
        \includegraphics[width=\textwidth]{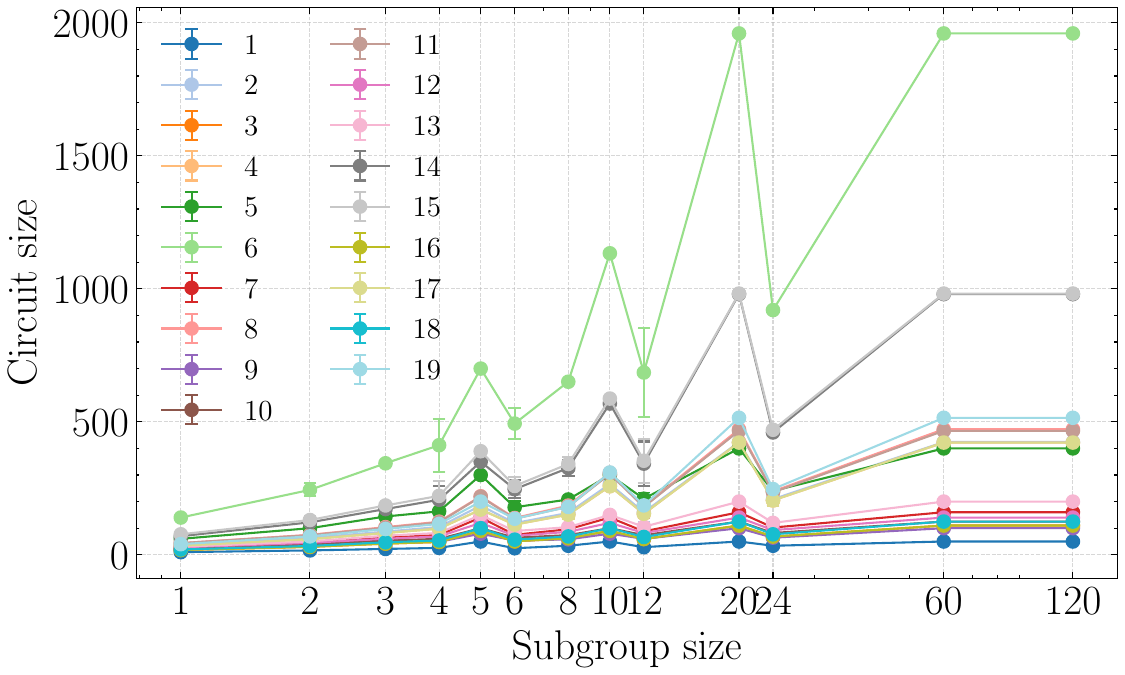}
        \caption{$S_5$ (five qubits, depth $1$).}
        \label{fig:absolute_circuit_sizes_5}
    \end{subfigure}
    \caption{Circuit overheads under subgroup-based symmetrization. We report the synthesized circuit size (total operations after Qiskit transpilation, opt-level~3) as symmetry increases (subgroup size on the x-axis). Most ansatzes grow moderately, while a few families incur sharp blow-ups and occasional non-monotonic changes across subgroup orders.}
    \vspace{0.5em}
    \label{fig:combined_circuit_sizes}
    \Description{A figure containing two line plots (Fig. 5a and Fig. 5b) comparing the absolute circuit size (Y-axis) versus the subgroup size (X-axis) of the symmetric group for 19 different Ansatz IDs (1 through 19). Fig. 5a shows the comparison for four-qubit ansatzes across subgroups in S4 (size up to 24). Fig. 5b shows the comparison for five-qubit ansatzes across subgroups in S5 (size up to 120). Both plots demonstrate that the circuit size, representing the number of operations, generally increases with larger subgroup sizes, with certain ansatzes (e.g., Ansatz 11 and 16) showing significantly higher complexity. Error bars are included for all data points.}
\end{figure*}

We start by comparing the sizes of the original circuit to the symmetrized circuits. The circuit size is the total number of instructions in the circuit. The sizes of depth-$1$ ansatzes for four qubits are presented in Fig.~\ref{fig:absolute_circuit_sizes}, and for five qubits in Fig.~\ref{fig:absolute_circuit_sizes_5}.

In \F\ref{fig:absolute_circuit_sizes}, most ansatzes grow moderately as symmetry increases, remaining below $200$ operations even at $|S'_k|{=}24$. However, a small number of circuits dominate the cost tail. The most expensive circuit to implement is the circuit with id $6$. While symmetrization creates substantial overhead for a couple of circuits, in many cases, the circuits do not become so long that the hardware cannot support them. Interestingly, in all cases, the cost of implementing symmetric circuits for subgroups of sizes $12$ and $24$ is the same.

We also observe that gate counts do not grow monotonically. This is especially apparent in Fig.~\ref{fig:absolute_circuit_sizes_5}, where circuits corresponding to subgroups of size $6$, $12$, and $24$ show lower overhead compared to the cases around them. We believe that the reason for this is the same as the reason why we observed the lower operator norm for the size $6$ subgroup in Fig.~\ref{fig:commutator_norm_with_twirling}. A similar phenomenon can be seen in the five-qubit cases in Fig.~\ref{fig:absolute_circuit_sizes_5}. The figure shows peaks at the same subgroups as we observe higher norm in Fig.~\ref{fig:commutator_norm_with_twirling_5}.

\subsection{RQ3: How does symmetry affect expressibility and entangling capability?}

\begin{figure*}
    \centering
    \includegraphics[width=0.7\linewidth]{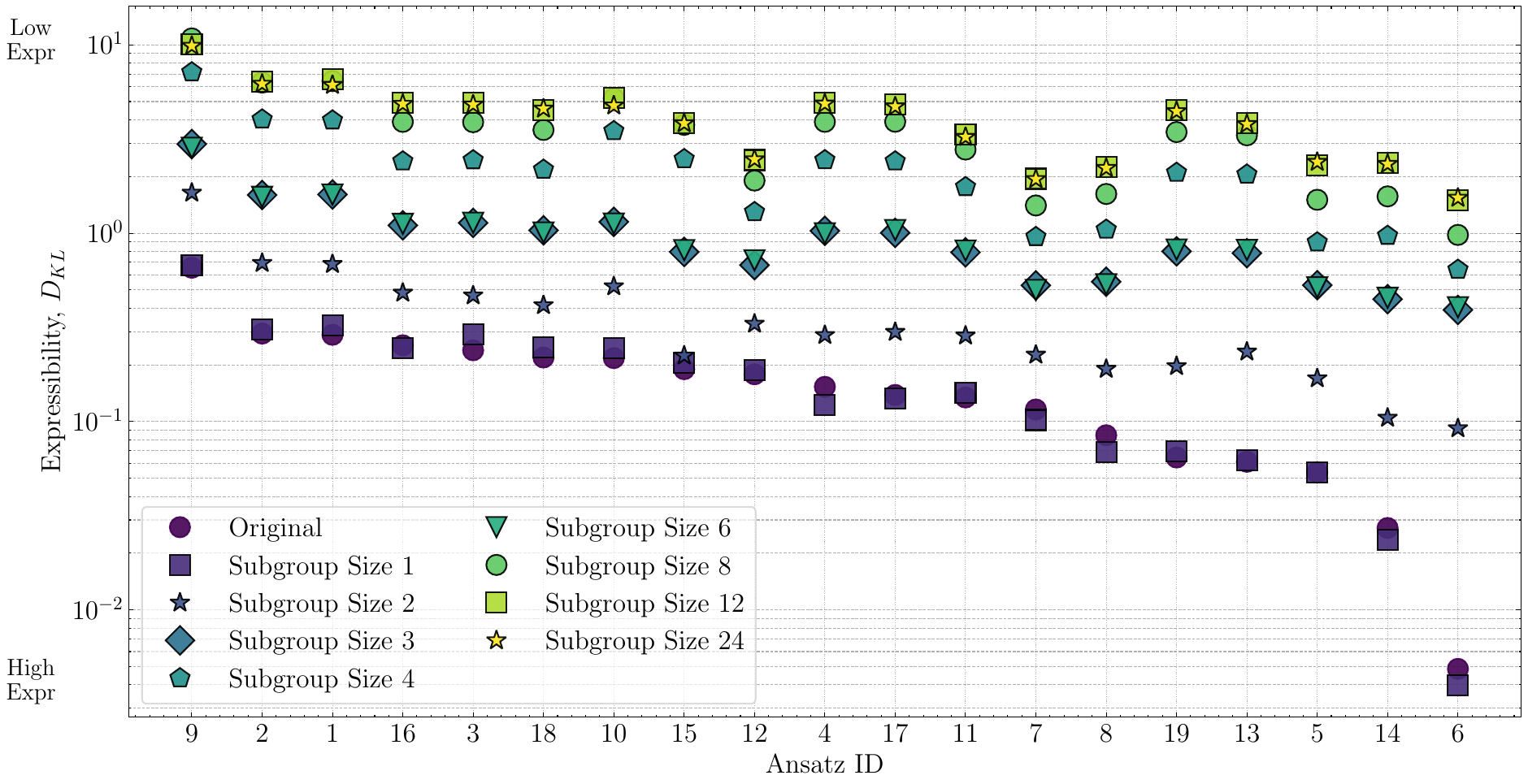}
    \caption{Expressibility under subgroup-based symmetrization (depth $1$): larger subgroups increase $D_{\mathrm{KL}}$ (lower expressibility).}
    \label{fig:expressibility}
    \Description{A scatter plot displaying the Expressibility (DKL, Y-axis) of 19 different ansatzes (labeled by Ansatz ID on the X-axis, ordered by decreasing original expressibility) for depth 1 circuits. The Y-axis is plotted on a logarithmic scale, with labels indicating Low Expressibility (top) and High Expressibility (bottom). Each Ansatz ID has multiple data points, representing the expressibility of the Original circuit (purple squares) and the circuits symmetrized using subgroups of size 1, 2, 3, 4, 6, 8, 12, and 24 (corresponding to S4). The plot primarily illustrates that increasing the size of the imposed symmetry subgroup (moving from purple squares to yellow crossed squares) generally decreases the expressibility of the resulting circuit.}
\end{figure*}

\tinyskip
\mypar{Expressibility}
To highlight differences in the sizes of the symmetry groups, we again fix the circuit depth to $1$ and plot the results as a function of subgroup sizes. Fig.~\ref{fig:expressibility} shows the expressibility results for the studied circuits. The results are ordered from the least to the most expressible, according to the expressibility value of the original circuit.

One can observe that the most expressive ansatzes are also the most expensive to implement, as shown in Fig.~\ref{fig:combined_circuit_sizes}. The values for the original ansatzes are close to those obtained in the original study~\cite{https://doi.org/10.1002/qute.201900070}. We consistently observe that expressibility decreases as symmetry increases. In most cases, the relative decrease in expressibility stays approximately the same.

Furthermore, we see that the expressibility of symmetric ansatzes corresponding to subgroups of sizes $3$ and $6$ yields similar KL-divergence values. We assume that this phenomenon again follows from the reasons that we discussed around Fig.~\ref{fig:commutator_norm_with_twirling} and Fig.~\ref{fig:commutator_norm_with_twirling_5}. Subgroups of size $3$ are often subgroups for subgroups of size $6$, and thus they show similar performance. The KL-divergence values also seem to be the same for symmetric ansatzes corresponding to subgroups of sizes $12$ and $24$, presumably for the same reason.

\begin{figure*}[t]
	\centering
    \begin{subfigure}[t]{0.4\textwidth}
    	\includegraphics[width=\textwidth]{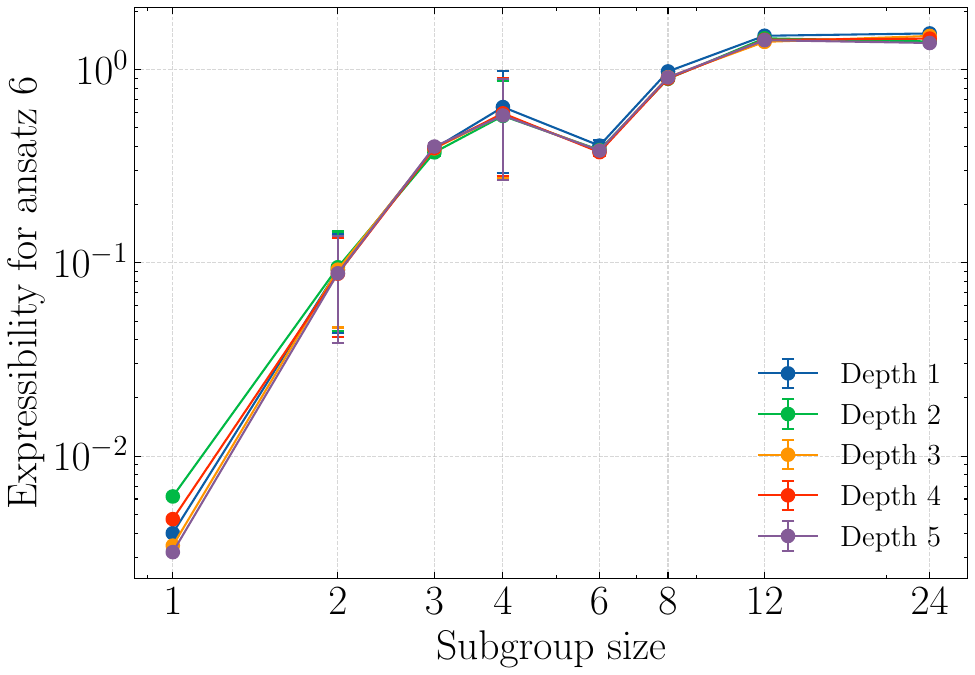}
    	\caption{Ansatz~6 expressibility: subgroup effects persist across depths.}
    	\label{fig:layerwise_expressibility_6}
    \end{subfigure}
    \hfill
    \begin{subfigure}[t]{0.4\textwidth}
    	\includegraphics[width=\textwidth]{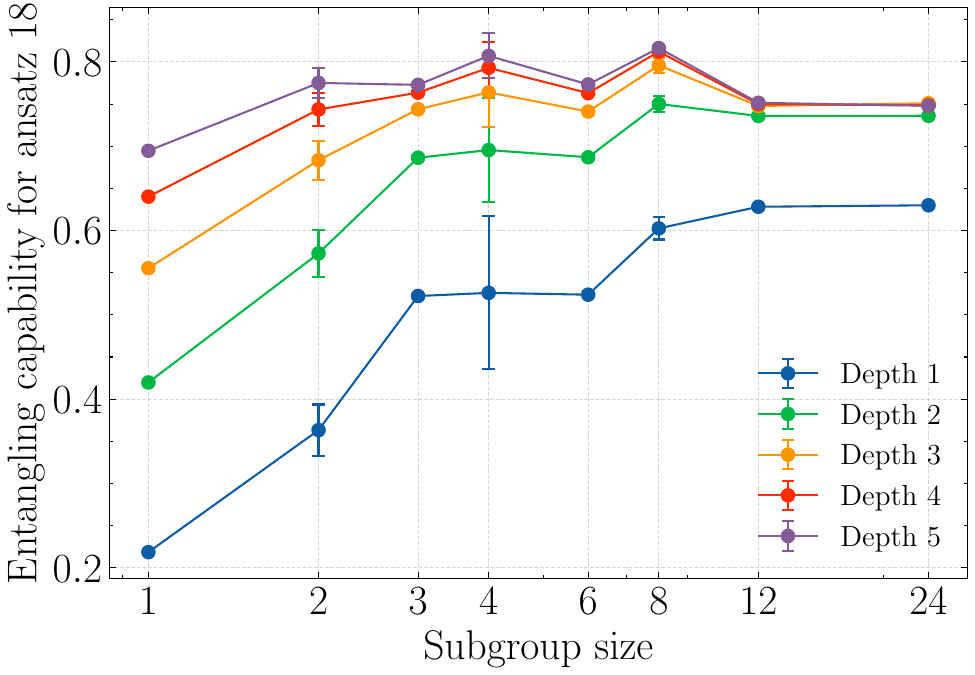}
    	\caption{Ansatz~18 entanglement: depth raises entanglement capability at weak symmetry and saturates under strong symmetry.}
    	\label{fig:layerwise_entanglement_capability_18}
    \end{subfigure}
    \caption{Depth sensitivity under subgroup-based symmetrization: depth shifts absolute values, while relative subgroup trends remain stable.}
	\label{fig:layerwise_sensitivity}
	\Description{Two line plots reporting how expressibility (left) and entangling capability (right) vary with subgroup size and circuit depth for representative ansatzes. The trends across subgroup sizes remain consistent as depth increases, suggesting that depth primarily changes absolute values rather than the relative symmetry effect.}
\end{figure*}

In the studied cases, the expressibility results indicate that most of the symmetrized circuits are not in the previously identified favorable expressibility region~\cite {https://doi.org/10.1002/qute.201900070}. Fig.~\ref{fig:layerwise_expressibility_6} demonstrates that even increasing the depth does not seem to increased reduced expressibility, and this applies to the other circuits in the study. Thus, the apparent solution is to use only sufficiently small or partial symmetries. It is possible that analyzing symmetries in the learning problem becomes part of the QML architecture design, allowing us to find the optimal model design.

\begin{figure*}
    \centering
    \includegraphics[width=0.7\linewidth]{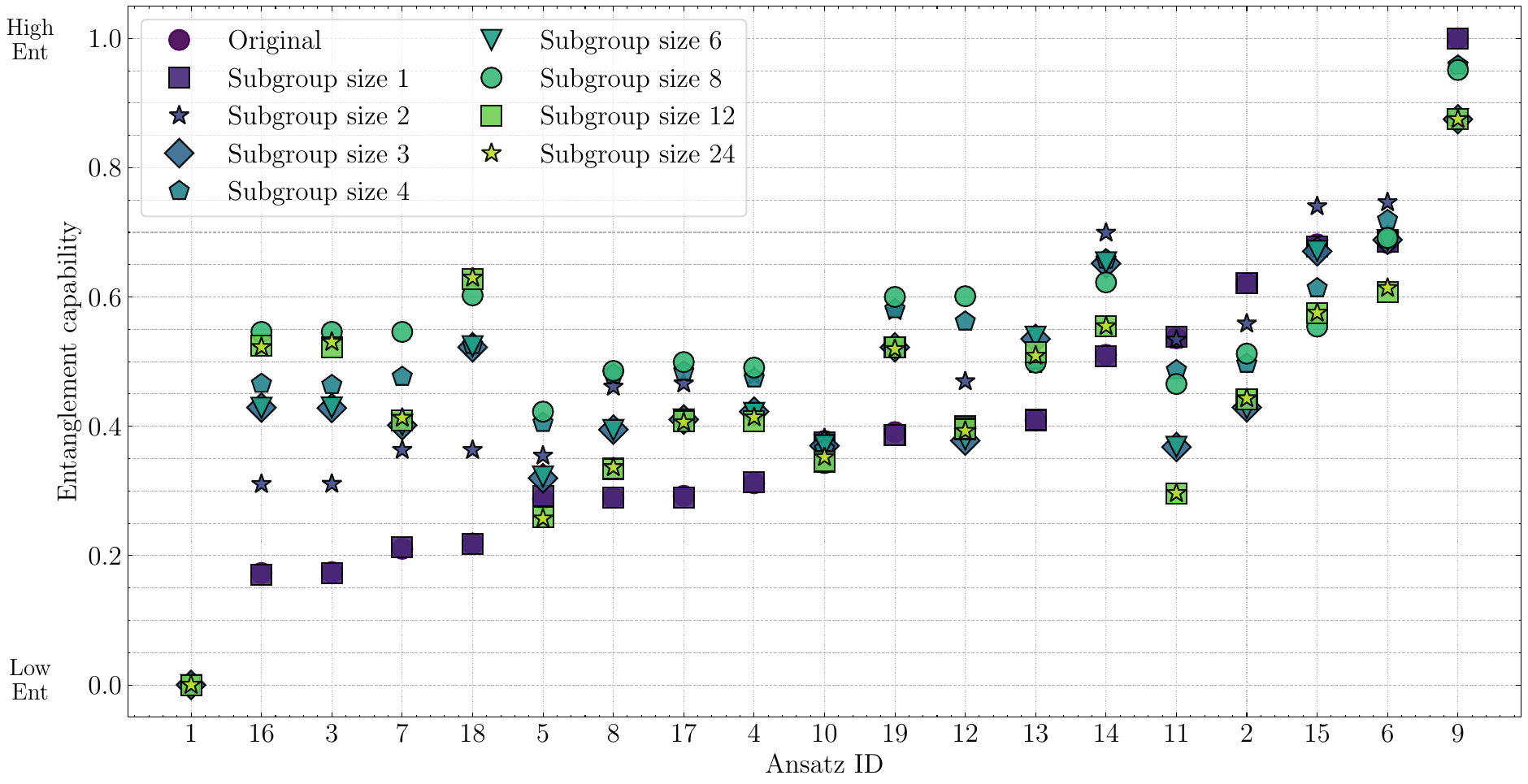}
    \caption{Entanglement capability of the ansatzes when depth is $1$}
    \label{fig:entanglement}
    \vspace{1em}
    \Description{A scatter plot displaying the Entanglement Capability (Y-axis) of 19 different ansatzes (labeled by Ansatz ID on the X-axis) for depth 1 circuits. The Y-axis is linear, ranging from 0.0 to 1.0, with labels indicating Low Entanglement (bottom) and High Entanglement (top). Each Ansatz ID has multiple data points, representing the entanglement capability of the Original circuit (purple circles) and the circuits symmetrized using subgroups of size 1, 2, 3, 4, 6, 8, 12, and 24 (corresponding to S4). The figure generally indicates that symmetrization increases entanglement capability compared to the original circuit for most ansatzes, as the colored subgroup points are typically positioned higher than the original purple points.}
\end{figure*}

\tinyskip
\mypar{Entanglement capability}
Fig.~\ref{fig:entanglement} presents the results of the entangling capability studies. While it is intuitive that expressibility decreases as symmetry increases, it is less intuitive how entanglement behaves in this context. We did not observe similarly consistent patterns as in the case of expressibility.

The general trend that we observe is that entanglement increases in most cases. This is most likely because the symmetrization introduces parametrized gates that create entanglement. Interesting exceptions are the ansatzes $2$, $9$, $10$, $11$, and $15$ that have non-parametric two-qubit gates. The results indicate that these non-parametrized entangling gates are transformed into gates that create less entanglement in the circuit (ansatzes $2$, $9$, $11$, $15$) or that maintain approximately the same entanglement capability (ansatz $10$). We also note that ansatzes $3$ and $16$ exhibit very similar characteristics, which is understandable given their structures.

None of the previous metrics examined the effect of increasing the circuit depth. We found that although increasing the circuit depth changes the absolute values, the relative influence of symmetries is consistent across depths for each ansatz. In essence, the patterns observed at depth $1$ already reflect the main effects of symmetrization, and greater depths rarely alter this picture. For example, present the expressibility values for ansatz $6$ with different depths in Fig.~\ref{fig:layerwise_expressibility_6}, which shows that the effects of symmetrization affect similarly. In the same way, Fig.~\ref{fig:layerwise_entanglement_capability_18} shows the entangling capability results for ansatz $18$.

\section{Discussion}



Across 19 ansatz patterns, subgroup-based symmetrization exposes a consistent symmetry--overhead--performance trade-off across ansatzes. As subgroup size increases, (i) generator drift grows, (ii) synthesized circuits become larger, and (iii) expressibility decreases, while entangling capability often increases. Without symmetrization (subgroup size $1$), our evaluation reproduced the reference baselines reported in~\cite{https://doi.org/10.1002/qute.201900070}, confirming that \sys preserves the original benchmark.

\tinyskip
\mypar{Why expressibility drops under stronger symmetry}
Large subgroups cause the largest decrease in expressibility because they enforce invariance across many qubits simultaneously. Learning problems that exhibit all these symmetries are often relatively constrained and structured. We still need to determine the exact practical consequences, because the low expressibility of symmetric models should still perform well on data that share the same symmetries. Thus, whether low or high expressibility and entangling capability values are desirable depends on the specific use case and the underlying hardware constraints.

\tinyskip
\mypar{Implications for quantum software engineering}
This work and the implementation open possibilities for tooling and automation for creating and analyzing symmetry effects in QML. The pipeline demonstrates that properties such as expressibility, entangling capability, and symmetry-induced overhead can be automatically estimated. \sys and similar tools could be used in many ways, such as ranking ansatz choices, deciding whether to adjust the degree of symmetry in ansatzes, stopping training early, or switching between ansatzes. When implementing such tools, one should keep in mind that expressibility and entangling capability are proxy metrics, and the models' true performance will depend on many factors. 

\tinyskip
\mypar{Scope and limitations}
This was not a study of real-world use cases, and thus, we did not examine how symmetries affect trainability in the setup we proposed in this work. This will be part of future work. Another promising direction is to investigate the effect of symmetries on measurement operations, which have been identified as less studied than other parts of quantum machine learning models~\cite{uotila2025perspectives}. 

We have also excluded studies on more complex data encoding methods, such as amplitude encoding, which should be included in future studies. Computing an induced unitary representation for amplitude encoding is more challenging. It might be easier to compute a specific relaxation to induced unitary representations defined in \autoref{eq:induced_unitary_representation}. If we fix the standard reference state $|00\cdots0\rangle$, we obtain $U_{\mathrm{init}}(\varphi(s, x))|00\cdots0\rangle = U_s U_{\mathrm{init}}(x) U_s^{\dag}|00\cdots0\rangle$. We call these state-level induced unitary representations because they are computed with respect to fixed states. Then, we compute $U_s$ only so that the previous state-level equation holds. This creates an induced unitary representation that does not necessarily satisfy the operator-level \autoref{eq:induced_unitary_representation}. 

\section{Conclusion} \label{sec:conclusion}
This paper presented \sys, an automated symmetry-aware QML pipeline that operationalizes subgroup-based symmetrization as a practical knob between no symmetry and full-group symmetry. Across 19 common ansatz patterns, \sys quantifies how increasing symmetry reshapes generator structure, circuit overhead, expressibility, and entangling capability, revealing a consistent trade-off: larger subgroups create higher overhead, reduce expressibility, and often increase entanglement. \sys and the results help practitioners reason about deployability and the expected performance of ansatzes early in the design of QML models. Our current scope focuses on permutation symmetries and angle encoding; future work includes extending \sys to richer symmetry groups and representations, supporting more complex encodings and measurements, and validating the trade-offs on end-to-end tasks and real-world datasets.

\begin{acks}
This work is funded by Business Finland (grant number 169/31/2024), Research Council of Finland (grant number 362729) and the \emph{Finnish Quantum Flagship Exploratory Project} to PI Bo Zhao. 
\end{acks}

\balance
\bibliographystyle{ACM-Reference-Format}
\bibliography{ref}

\end{document}